\documentclass[aps,prd,twocolumn,longbibliography,nofootinbib]{revtex4-1}
\usepackage{amsmath,amsfonts}

\begin{document}

\newcommand\be{\begin{equation}}
\newcommand\ee{\end{equation}}

\title{Why Gauge?}

\author{Carlo Rovelli}
\affiliation{Aix Marseille Universit\'e, CNRS, CPT, UMR 7332, 13288 Marseille, France.\\ Universit\'e de Toulon, CNRS, CPT, UMR 7332, 83957 La Garde, France.}
\date{\today}

\begin{abstract}\noindent
The world appears to be well described by gauge theories; why? I suggest that gauge is more than mathematical redundancy. Gauge-dependent quantities can not be predicted, but there is a sense in which they can   be measured. They describe ``handles" though which systems couple: they represent real relational structures to which the experimentalist has access in measurement by supplying one of the relata in the measurement procedure itself.   This observation leads to a physical interpretation for the ubiquity of gauge: it is a consequence of a relational structure of physical quantities.
\end{abstract}

\maketitle

\section{Introduction}
The year 2013 has been marked by two pieces of notable news from Nature: the CERN study of the Higgs field \cite{Aad20121,Chatrchyan201230} and the Planck satellite's observations of the cosmic background radiation \cite{Collaboration:2013fk}. Both support, far beyond the expectations of many theorists, the effectiveness of the current description of the world, based on the standard model of particle physics and general relativity.  These two theories are, each in its own way, gauge theories.  This was pointed out by Utiyama \cite{Utiyama:1956sy} shortly after the writing of the Yang-Mills equations and was formalized by Dirac \cite{Dirac:1950pj}.   Why is the world so well described by gauge theories?\footnote{For an interesting philosophical discussion of the problem, and general references, see \cite{Healey2007,Rickles:2008fk}.}   

Gauge theories are characterized by a local invariance, which is often described as mathematical redundancy. According to this interpretation,  physics is coded into the gauge-invariant aspects of the mathematics.  This interpretation obviously captures something important. Dirac defines gauge as under-determination of the lagrangian variables' evolution, and observes that if  evolution is deterministic, as we expect in the classical theory, then only gauge-invariant quantities can be physical, by definition.\footnote{This interpretation has gained support from recent  results on the possibility of computing amplitudes using methods that make gauge look just like a complication \cite{ArkaniHamed:2008gz}, and by AdS/CFT suggestions that gravitational physics can be coded on the asymptotic boundary. In the quantum-gravity community, it grounds the common idea that the space of solutions of the hamiltonian constraint  exhausts the physics of a system.}

But things are not so clear. If gauge is only mathematical redundancy, why the common emphasis on the importance of gauge symmetry? Why the idea that this is a major discovery and guiding principle for understanding particle physics? How did Einstein's intuition about covariance lead him to  unraveling relativistic gravity?

In this paper, I observe that the interpretation of gauge as pure mathematical redundancy disregards an aspect of the physics.  This aspect explains the effectiveness of gauge theory, and helps us unravel the actual meaning of the fact that the world is described by gauge theories. 

The basic observation is that although we can interpret a given physical system in terms of its gauge-invariant observables, we nevertheless \emph{couple} this same system with other systems using its non-gauge-invariant variables.\footnote{A related argument is Feynman's suggestion (\cite{Feynman:2011fk} II.15.3-4) that we should take the Maxwell potential (a gauge-dependent quantity) seriously on the ground that it determines the energy of a current (it appears in a coupling).} 
  
I develop this observation and consider a consequent general interpretation of the ubiquity of gauge in our world.  This leads to a different view of gauge-dependent quantities, indicating that, in an appropriate sense, they can be taken as unpredictable, but measurable.  

\section{Coupling gauge systems}
 
 A physical system can be considered in isolation or in the way it couples with other systems. 

Electromagnetism can be expressed as a $U(1)$ gauge theory for the Maxwell potential $A$.\footnote{Notation in this note is standard and can be found in all common theoretical-physics textbooks.}  The gauge-invariant content of the theory, whose evolution is predicted the Maxwell equations, is captured by the electromagnetic field $F=dA$.\footnote{Plus additional global quantities if the topology is non-trivial.} This  leads us to say that what exists in Nature is only the gauge invariant content of $A$. However, we describe the \emph{coupling} of the electromagnetic field with a fermion field $\psi$ in terms of the interaction lagrangian density 
\be
               L=\bar\psi(\gamma^\mu A_\mu)\psi 
               \label{Le}
\ee
written in terms of $A$, not $F$.   Similarly, we describe the gravitational field with the metric tensor $g$, which transforms under the gauge transformations of general relativity, namely diffeomorphisms. But the matter energy-momentum tensor couples directly to $g$,  
\be 
    L= \sqrt{-g}\, g_{\mu\nu}\, T^{\mu\nu}, 
               \label{Lg}
\ee
and not to a gauge-invariant function of $g$.

The  lagrangian terms \eqref{Le} and \eqref{Lg} are gauge-invariant.   But they are so in the \emph{coupled} systems, formed by $(A,\psi)$ and $(g,matter)$ respectively, where the gauge transformations act on all variables. The reason they are gauge-invariant is not because they couple gauge-invariant quantities. They are gauge-invariant couplings between gauge-\emph{variant} quantities. 

If we consider the electromagnetic or gravitational fields alone, \emph{or}  the system formed by coupling these to the Dirac field or matter, then we can view gauge quantities as a redundancy of the formalism.  But if we think in physical terms, we face an obvious ambiguity: if the \emph{real} electromagnetic field is purely described by $F$ alone, how can it couple to $\psi$?  If the gravitational field is  an equivalence class of metrics under diffeos, how can this couple locally to a particle?  Where, so to say,  does $A$ go when no fermion is around? 

These considerations suggest that gauge non-invariant quantities (``gauge variables", from now on) represent, in a sense, handles though which systems can couple.

\section{A model}

In order to get some direct intuition about the nature of gauge, consider a set of $N$  variables $x_n(t)$, with $n=1...N$ whose dynamics is governed by the lagrangian
\begin{equation}
       L_x = \frac12 \sum_{n=1}^{N-1} \left(\dot x_{n+1} - \dot x_{n}\right)^2. 
       \label{Lx}
\end{equation}
 where $\dot x_n\equiv dx_n/dt$. (A potential term depending on differences could be added without affecting the following discussion.) The equations of motion are invariant under the gauge transformation
\begin{equation}
                       x_n\to x'_n= x_n+\lambda
                       \label{gaugex}
\end{equation}
for an arbitrary function $\lambda(t)$.  This is a minimal prototype of a gauge system.  The evolution of $x_n(t)$ is under-determined by the equations of motion.  A complete set of gauge-invariant quantities is given by  
\begin{equation}
                       a_n= x_{n+1} - x_{n}, \ \ \ \ \ n=1...N-1,
                                              \label{da}
\end{equation}
whose evolution is deterministic. According to a standard interpretation of this system, the lagrangian \eqref{Lx}  describes the variables \eqref{da} \emph{and nothing else}; a gauge-invariant description of the same system is given by  the lagrangian
\begin{equation}
       \tilde L_x = \frac12 \sum_{n=1}^{N-1} \dot a^2_n  
\end{equation}
and the systems $L_x$ and $\tilde L_x$ are  equivalent. Are they really?

Consider a \emph{second} set of variables, $y_n(t)$ with $n=1...M$ whose dynamics
is similarly governed by the lagrangian
\begin{equation}
       L_y = \frac12 \sum_{n=1}^{M-1} \left(\dot y_{n+1} - \dot y_{n}\right)^2. 
\label{Ly}
\end{equation}
The equations of motion are invariant under the gauge transformation 
\begin{equation}
                       y_n\to y'_n= y_n+\lambda'
                       \label{gaugey}
\end{equation}
for an arbitrary function $\lambda'(t)$. 
As before, we can define an equivalent $\tilde L_y$ system just in terms of the gauge-invariant quantities $b_n=y_{n+1}-y_n$. 

Now, consider the \emph{coupled} system defined by the variables $x_n, y_n$ and the lagrangian
\begin{equation}
       L = L_x+L_y + L_{\rm int} 
       \label{Lt}
\end{equation}
where
\begin{equation}
      L_{\rm int} = \frac12 \left(\dot y_{1} -\dot  x_{N}\right)^2. 
       \label{Lint}
\end{equation}
The coupled system is invariant under the gauge transformations (\ref{gaugex}) and (\ref{gaugey}) if $\lambda(t)=\lambda'(t)$.   Its gauge-invariant observables are \emph{more} than those of the individual systems: they are given by gauge-invariant observables of the first system, those of the second, {\em plus a new one:} 
\begin{equation}
                       c= y_{1} - x_{N} , 
                       \label{dnew}
\end{equation}
which depends on \emph{gauge} variables of both.  

Suppose we take the point of view that gauge is nothing else than mathematical redundancy.  If so, we should be able to describe the above coupling of the two systems $L_x$ and $L_y$ by directly coupling  $\tilde L_x$ and $\tilde L_y$, without  need of adding additional variables.  But this is not possible.  We cannot obtain the coupled system  (\ref{Lt}) by coupling the gauge-invariant systems $\tilde L_x$ and $\tilde L_y$, because the number of variables of these are $(N-1)+(M-1)=N+M-2$, while the degrees of freedom of the coupled system are $N+M-1$, that is, one more.

We can obtain $L$ from $\tilde L_x$ and $\tilde L_y$ only by adding one extra variable by hand. But this variable was indeed already contained in the union of the $L_x$ and $L_y$ gauge variables. This shows how gauge variables play a direct physical role, when coupling systems. The system  $\tilde L_x$ is not equivalent to the system $L_x$ because the second codes  information about possible couplings.  

Thus, there is a precise sense in which  $\tilde L_x$ and $\tilde L_y$ are {\rm not} equivalent to  $L_x$ and $L_y$: the equivalence holds only as long as we consider a systems in isolation; it fails when the possibility of coupling  is considered. 

\section{Relative observables}

The variable $c$ defined in  \eqref{dnew} is a ``relative observable", in the sense that it is not a function of the sole variables of one single system.  It captures --so to say-- something about the value of a gauge variable of one of the two systems, $y_1$, ``relative" to the value of one gauge variable of the other, $x_N$.\footnote{In fact, all the variables of the system \eqref{Lt} are of the same form.  The system can be rewritten in the form 
\begin{equation}
       L = \frac12 \sum_{n=1}^{N+M-1} \left(\dot z_{n+1} - \dot z_{n}\right). 
       \label{L}
\end{equation}
where $x_n=z_n$ and $y_n=z_{n+N}$, which in turn can be split into two subsystems at any value of $n$, showing that any of its gauge invariant variables 
\begin{equation}
                       d_n= z_{n+1} - z_{n}  \hspace{3em} n=1, ..., N+M-1.
                       \label{d}
\end{equation}
is a relative observable, in the sense above.  Thus, \emph{all} gauge-invariant degrees of freedom of this system have such a relational character. }

It is enlightening to make the example above more tangible by imagining a concrete situation it could describe. Imagine a simple world where there are only $N$ spaceships capable of measuring their relative distances.  The position $x_n(t)$ of one individual spaceship cannot be determined in absolute terms, because there is nothing else with respect to which it can be determined. This is why the equations of motion leave it free.  The \emph{relative} positions can be measured and they obey the equations of motions determined by the theory \eqref{Lx}. 

Now imagine that a  fleet of alien starships approaches from far away, described by the theory \eqref{Ly}.  As the two fleets starts communicating,  one new observable appears: the position of $y_1$ with respect to $x_N$.  

In the example, gauge-invariant degrees of freedom are relational observables, relating variables that  in the absence of coupling are gauge degrees of freedom. 

The gauge invariance of $L_x$ is invariance under an arbitrary time-dependent displacement of all spaceships. The position of the individual ships is redundant in the theory insofar as we measure only relative distances among these.  But in the physical world each ship has a position nevertheless:  this becomes meaningful with respect to one additional ship, as this appears. In other words, the existence of a gauge expresses the fact that a reference is needed to measure the position of a ship.  It expresses the fact that we measure position relationally. 

I suggests that this is the general case for all gauge systems.

The fact that the world is well described by gauge theories expresses the fact that the quantities we deal with in the world are generally quantities that pertain to \emph{relations} between different parts of the world, that is, which are defined across subsystems.\footnote{Of course the Yang-Mills connection connects the internal frame orientations of \emph{two} points of spacetime and the metric expresses the distance between \emph{two} points of spacetime.} 

The example shows how a gauge quantity typically describes an individual component of a relative observables. 

\subsection{General relativity}

Let me now apply the idea illustrated above to the real world.
General relativity is invariant under the transformation
\begin{equation}
      g_{\mu\nu}(x)\to g'_{\mu\nu}(x)
      =\frac{\partial f^\rho(x)}{\partial x^\mu}\frac{\partial f^\sigma(x)}{\partial x^\nu}g_{\rho\sigma}(f^{-1}(x)).
\end{equation}
of the metric tensor, which is  a gauge transformation in the sense of Dirac, 
and implies under-determination of the evolution of the metric under Einstein equations.  Gauge-invariant quantities in pure GR are invariant under the transformation above. 
A function of the metric such as   
\begin{equation} 
\tau = \int_0^1 d\tau \sqrt{g_{00}(\tau,0,0,0)}. 
\label{dgr}
\end{equation}
 fails to be gauge-invariant. 

Consider the system formed by GR and two particles 
described by the worldlines $x^\mu(\tau)$ and $y^\mu(\tau)$. These 
transform under gauge transformations as 
\begin{equation} 
      x^\mu(\tau)\to x'{}^\mu(\tau)= f^\mu(X(\tau)). 
\end{equation}
and similarly for $y^\mu(\tau)$.  Suppose for simplicity that we consider only solutions 
where the two worldlines meet at two points $P$ and $Q$.   Let 
$T$ be the proper time from $P$ to $Q$ along the worldline of the first particle.  
\begin{equation} 
T = \int_P^Q d\tau\  \sqrt{g_{\mu\nu}(x(\tau))\frac{dx^\mu(\tau)}{d\tau}\frac{dx^\nu(\tau)}{d\tau}}. 
\label{distance}
\end{equation}
The quantity $T$ is a gauge-invariant quantity of the coupled
system formed by pure general relativity and the two particles. 

In fact, it can be measured and predicted: if the first of the two particles is the Sun and second is the  
Halley comet, then $T\sim 75 \ years$, as was indeed predicted by Edmond Halley.

Does this observable $T$ pertain to the gravitational field or to the particles?
The answer is clearly to both.  It is a relational observable in the sense above.
It is the analog of the observable $c$ in \eqref{dnew}.
It measures a line integral of the gravitational field along a worldline determined
by a particle. Neither the position of the participle nor this line integral in the
absence of the particles are gauge-invariant.

\subsection{Yang-Mills}

The physical meaning of the gauge invariance of general relativity is easier
to understand than that of Yang-Mills
theory.  The gauge invariance of general relativity is the implementation of a
physical idea:  
localization is entirely relative (hence the name ``general relativity"). Localization 
is always determined with respect to something
else, in particular for instance with respect to the gravitational field. 
We  understand the gauge invariance of the theory as a realization of the
idea of relative observables: the position of the particle in the above example  
makes sense only in relation to the gravitational field, and the value of the gravitational 
field at a point makes sense only in so far as there is something physical with respect to 
which that point is defined.  This is the bread and butter of every general relativist. 

The physical meaning of the Yang-Mills gauge invariance, I think,
can be understood similarly.  To illustrate it, the most transparent example is 
the original $SU(2)$ theory of Yang and Mills \cite{Yang:1954ek}, with two fermions
$p$ and $n$ (erroneously identified with proton and neutron at the time) in a multiplet.
The local $SU(2)$ gauge symmetry of the theory transforms the non-abelian
gauge field $A$ as well as the fermion's fields.  

Suppose 
we observe one fermion at a point $x$.  Can we say if it is a $p$ particle or an $n$
particle?  The answer of course is subtle.  A priori, there is nothing that 
distinguishes the two, so, observing one particle we can arbitrarily call it
$p$ or $n$ as we wish.  But once one particle has been named, any
other particle at a point $y$  connected by a given path $\gamma$ to the 
first is uniquely determined as a given combination of $p$ and $n$. This is, of course, because the gauge field
defines a parallel transport operator along the path, which allows for the 
angle between the two particles' states $\psi_x$ and $\psi_y$ in the internal 
space to be well defined with respect to one another. 
In other words, the quantity 
\be
\theta = \langle \psi_x|Pe^{\int_\gamma A}|\psi_y\rangle
\label{theta}
\ee
is gauge invariant. Here we see again how a gauge invariant quantity
is a relational quantity, relating gauge field variables and two gauge fermion 
variables.  The nature of a particle, whether it is $p$ or $n$, is something
which is defined in relation to the field and other particles, in the same
manner in which a direction in spacetime is only defined with respect
to the gravitational field and something else. 

These relations are cleanly expressed in the geometrical 
representation of Yang-Mills theory in terms of fiber bundles \cite{Eguchi:1980jx}. 
In this geometrical picture, the connection is a field that 
establishes a preferred \emph{relation} between a frame
in a fiber and a frame in a nearby fiber. In other words, it
is a quintessential relational observable. Gauge invariance
is then the freedom of choosing a frame in each fiber. 

This geometrical language emphasizes the similarity between 
Yang-Mills theory and general relativity.  In both cases 
gauge invariance is a matter of arbitrary coordinates. 
As I emphasize below, in both cases the choice of a
particular gauge can be realized \emph{physically} via 
a coupling: with a material reference systemic general
relativity, with the choice of reference fermions in Yang Mills
theory. Therefore the geometry of the bundle describes 
the possibility of coupling for the gauge field. 

Gauge theories are sometime introduced mentioning the
historical idea of promoting a global symmetry to a local
one. The purpose of the field would be to realize the local 
symmetry.\footnote{I have always found this story a bit 
mysterious. Entities in Nature are not there to realize  
purposes: they are there and have properties. The 
Yang-Mills field is there and has symmetries; it is not there 
for the purpose of implementing a symmetry. We should
distinguish our understanding of Nature from  the path of 
discovery. If we understand that large wild mammals have
disappeared from America because of human hunting,
we should not conclude that humans exists for the purpose
of exterminating mammals.}  This idea, however, leaves
the question I am addressing in this note open: 
why do we need to describe the world with local symmetries if 
we then interpret these symmetries as mathematical redundancy?

\subsection{Time as gauge}

The last example of gauge quantity I examine, before discussing observability of these quantities in general, is that generated by the hamiltonian constraint of a covariant theory  \cite{Dirac:1950pj}.
It is a well-known observation that
any standard hamiltonian system can be reformulated as a gauge system with vanishing 
hamiltonian.   A conventional Newtonian system with $N$ variables $q_n(t)$ and lagrangian
\begin{equation}
       L(q_n,\dot q_n) =  \sum_{n=1}^{N} \frac{\dot q_n^2}{2}-V(q_n). 
\end{equation}
is equivalently described by the $N+1$ variables $(t(\tau),q_n(\tau))$ and lagrangian
\begin{equation}
       \tilde L(t,q_n,\dot t,\dot q_n) =  \sum_{n=1}^{N} \  \frac{\dot q_n^2}{2\dot t}-\dot t \, V(q_n). 
       \label{Lxx}
\end{equation}
where now the dot indicates the derivative with respect to a new lagrangian evolution parameter $\tau$. 
 This theory is invariant under the gauge transformations
 \begin{eqnarray}
                       q_n(\tau)\to q'_n(\tau)&=& q_n(\lambda(\tau)) \\
                          t(\tau)\to t'(\tau)\   &=& t(\lambda(\tau))
                       \label{gaugexx2}
\end{eqnarray}
for an arbitrary function $\lambda(\tau)$.  The physical interpretation of the variable $t$ is transparent: it is simply the time $t$ of the original theory (hence the same name), promoted to a full fledged lagrangian variable $t(\tau)$, in a theory where the new lagrangian evolution parameter $\tau$  is now a pure-gauge quantity. 

The relational aspect of the gauge is manifest in this case: the gauge-invariant content of this system is in the \emph{relative} evolution of $q_n$ and $t$. The quantities $q_n(\tau)$ and $t(\tau)$, taken individually, are gauge variables.  Evolution in the lagrangian evolution-parameter $\tau$ is pure gauge. The hamiltonian turns out to be weakly vanishing. The physical evolution is coded in the \emph{relative} evolution of all lagrangian variables.\footnote{Physics is about  relations and this is  implemented in the standard dynamics, which codes relations between  values of physical variables --including $t$-- or, equivalently, the evolution of all variables in a single variable ($t$) taken as reference. A detailed discussion of this point is in Section 3.2.4 of \cite{Rovelli:2004fk}.} 

Notice that we are used to considering the quantities $q_n$ as observables, but also to assuming that the time $t$ can be determined by simply looking at a clock.  But here the clock plays precisely the role of the position of the alien starship above. That is, the temporal evolution of the quantity $q_n$ is well defined \emph{only in so far as we can measure it with respect to a clock variable $t$}. The gauge-invariant observable is not $q_n$, it is $q_n(t)$ at some observed time $t$.   The quantities $q_n$ and $t$, taken individually, have no determined dynamics.  

The indeterminacy associated to the gauge freedom is concretely realized by the fact that (while  the value of other variables at some time can be predicted) \emph{the value of the clock variable itself can not be predicted}.\footnote{This interpretation of the relation between gauge and time evolution is reinforced by the analysis of the coupling of general-covariant systems developed in \cite{noi}.  When coupling general-covariant systems, one finds systems of hamiltonian constraints generating multi-dimensional orbits. If the physics singles out a preferred time variable, the others independent variables appear as gauges. See \cite{noi}.}

\section{Observability of gauge variables}

I now come to a key point of this discussion: the measurability of gauge variables.  In physics, when we talk about measurement, we refer to an interaction between a measured system $S$ and a measuring apparatus $O$. 

For a system $S$ with gauges, nothing prevents us from measuring a property of a system $S$ by means of a (gauge-invariant) interaction between a gauge variable of $S$ and a gauge variable of $O$. 

When we do so, we can determine a number for a gauge variable of the system $S$. That is, we can ``measure" a gauge variable (below are several examples). But this number cannot be predicted by the dynamical equations of the system $S$. Quantities that can be measured but not predicted were denoted ``partial observables" in \cite{Rovelli:2001bz}. 

The conclusion that gauge-variables can be measured does not contradict Dirac's original argument according to which considering  a gauge variable as ``observable" leads to indeterminism. In fact, it confirms it.  Dirac used ``observable" in the sense of predictable quantity.  We can turn Dirac's argument around: every time that the physics is such that a certain measurable quantity entering the description of a system is unpredictable in terms of the system's dynamics alone (like the time, or the position of a single starship), it must be described  in the formalism by a gauge variable. This is what gauge variables are. 

This happens often in the real world.  For instance, a gravitational wave detector measures some 
components of the metric field.  By themselves these components are not gauge invariant, but 
they correspond to a gauge-invariant observable of the field+detector system.   

Again, in general relativity we can label the points along a worldline with the proper time $\tau$ from an initial event.  As long as no other event is singed out, $\tau$ is a gauge variable, because we do not know where we are along the worldline when we measure it.  But if we have a second similar gauge variable, the combination of the two gives a gauge-invariant quantity.   
For instance, say I keep a precise clock in my hands and throw upward a similar clock, which then falls down. When the two clocks meet again, the one in my hand will be late with respect to the one in free fall.  Given  initial data, the theory predicts the reading of the free falling clock $\tau(\tau')$ as a function of the time $\tau'$ of the clock in my hands. Equivalently, it gives $\tau'(\tau)$.   The two quantities $\tau$ and $\tau'$ are gauge variables, while the value of one at a given value of the other is a gauge-invariant quantity.  They are both  measurable. But only their relative value is predictable. 

A particularly clarifying example is provided by the local $SO(3,1)$ gauge invariance of the tetrad formulation of general relativity.  On the one hand, the local $SO(3,1)$ gauge transformations are mathematical redundancy, since (at least in the absence of fermions) the theory can equally be formulated in terms of tetrads or in terms of the metric. But on the other hand, the tetrad components have a transparent physical interpretation, since they can be viewed as axes of a concrete material physical system realized for instance by our laboratory.  The tetrad components are often concretely used in this manner in practical applications of the theory.  How can something be mathematical redundancy and at the same time concretely determined?   

The answer is that when we couple general relativity to the matter of a material reference system, the components of the gravitational field with respect to the directions defined by this system are gauge-invariant quantities of the coupled system; but they are gauge-dependent quantities of the gravitational field, measured with respect to a given external frame. 

Similarly, a detector that recognizes the $p$ particle in the original Yang Mills theory is in principle possible; what it is actually measuring is the angle in internal space between the measured particle and a $p$ standard in the detector itself.  

Suppose we have several  apparatus that measure whether particles are $p$ particles or $n$ particles, \emph{with respect to a local standard particle defined as $p$}.  In each there is a fermion, and a way to measure the (gauge-invariant) angle $\theta$ in internal space, given in \eqref{theta},  between an incoming particle and this reference fermion.  Say the apparatus are at the boundary of a process.  The result of \emph{one} measurement is unpredictable  because it depends on the (arbitrary, as far as the system's dynamics is concerned) state of one reference fermion in the apparatus.   Notice the similarity between this situation and the ships: the internal state of each fermion is determined only with respect to another one, like the position of the ships is determined with respect to another one.

Finally,  the prototypical example of a partial observable is time: a quantity that we routinely determine (looking at a clock) but we can not predict from the dynamics of the system. 

There is a common idea that one can restrict the formulation of a general covariant system with vanishing hamiltonian to its physical space $\Gamma_{ph}$, and ignore its partial observables, since they are non-gauge-invariant.  The problem with this idea is that if we do so, we  have then to recover missing structure in a convoluted manner. Take a harmonic oscillator  formulated as a re-parametrization invariant system on a two-dimensional extended configuration space with coordinates $(q,t)$ and Hamiltonian constraint $C=p_t+\frac12(p_q^2+q^2)=0$. The space  $\Gamma_{ph}$ is the space of the orbits of $C$ in the $C=0$ surface. It is a two-dimensional space, with coordinates $(A,\phi)$, and is blind to the physics of the oscillator.  This physics can be expressed by considering gauge-invariant observables such as the evolving constants of the motion $q_\tau=A \sin(\tau+\phi)$ labelled by a parameter $\tau$ \cite{RovelliQM3}; but this is a baroquism, which rephrases the fact that the physics is not in $\Gamma_{ph}$ alone:  it requires additional structure. In fact, it is in the relation between the points in $\Gamma_{ph}$ and the partial observables $q$ and $t$. The partial observables are relevant because we have measuring devices to determine them. The relevant solutions of the dynamical equations are, for each point $(A,\phi)$ in $\Gamma_{ph}$, relations between the partial observables $(q,t)$, such as
\be
f(q,t;A,\phi)=q-A\sin(t+\phi)=0.
\ee
(The full description of this formulation is for instance in Chapter  3 of \cite{Rovelli:2004fk}.) Ignoring the role of partial observables complicates the matter pointlessly.

Gauge quantities cannot be predicted, but can often be measured. They measure our  localization in time, our orientation in general relativistic space, in the internal space, and similar.\footnote{The fact that a gauge variable can be measured even though it cannot be predicted is not a paradox and it does not imply that a gauge theory is useless because it is unable to predict the results of possible measurements.  Any use of a physical theory, in fact, includes predictable quantities and unpredictable ones. The last can be identified with those determination the measurement location.  The main example is time: we look at the clock to read what time is it, and then we may predict the value of some variable at that time.  A theory can be useful and predictive even if it does not predict at which time we are going to look at the clock.}

\section{Boundary formalism}

We can formalize the observations in the previous section by using the covariant form of the dynamics \cite{Rovelli:2004fk} based on the  boundary formalism \cite{Oeckl:2003vu}.  Let us start with a simple Newtonian system with  degrees of freedom $q_n$ evolving in the time $t$.  We  measure $q_n$, the momenta $p_n$ and $t$ at initial and final boundaries of a process, and call $(q_n^{i},p_n^{i},t^{i})$ and $(q_n^{f},p_n^{f},t^{f})$ the outcomes. The dynamical equations determine constraints among these quantities. If the initial values are known,  the final values are not all determined: in particular, we can take $t^f$ to be unpredictable, and the other final quantities, $(q_n^{f},p_n^{f})$ to be determined as functions of $t^{f}$.

Now imagine that there is also a local $SU(2)$ gauge invariance relevant in the experiment, and let's disregard spacial dependence for simplicity.  Then in addition to the above data each degree of freedom has a component, 
say $\theta_n^i$ and  $\theta_n^f$ in internal space, which can be measured with respect to a given reference fermion at initial and final times.  Since the relative orientation of the reference fermions is not a priori known, {one of the} angles $\theta_n^f$ is not determined by the dynamics of the process. As far as the dynamics of the process is concerned, it is arbitrary.  Final variables are therefore functions of time \emph{and} this arbitrary angle.

Notice that the internal gauge dependence has the same formal structure as time. In the hamiltonian formalism, both can be expressed by a constraint: the  gauge-constraint for the internal gauge and the hamiltonian constraint for the time dependence.  In both cases, we can  associate measurement procedures to gauge variables at the boundary of a process.  These are realized by gauge-invariant interactions between a gauge variable of the system and a gauge variable of the apparatus. 

This shows that the the hamiltonian constraint and the other constraints can be treated on equal footing: in both cases, they determine dynamical constraints among non-gauge-invariant observables, or partial observables, measured at the boundaries of a process. 

In the canonical framework, the Yang-Mills constraint dictates how variables change with respect to a change of the internal boundary frame. The diffeomorphism constraint dictates how variables  change with respect to a change in the location of the spatial boundary reference frame. The hamiltonian constraint dictates how all variables collectively change with respect to a change in the temporal location of the boundary, in particular with respect to the variable used as clock.  These constraints code the full content of our dynamical theory. 

In all cases, indeterminacy in some variables is due to the arbitrariness of the choice of the internal frame, the spacial reference, or the time of the measurement. Dynamics is the study of relations between partial observables, which are, in general, gauge-dependent quantities of a system to which we can couple an apparatus.\footnote{This raises an intriguing question in the quantum context. In quantum theory we compute probabilities for outcomes when a system $S$ interacts with an ``apparatus" $O$. According to the preferred interpretation of quantum theory, we  view such interactions as a measurement by a classical apparatus (Copenhagen); a generic physical interaction (relational interpretation); the establishing of an entangled state with an external
system (many-world); possibly with effective suppression of interference terms (decoherence);
a physical event not yet described by standard theory (GRW, Penrose);  
or an interaction revealing the underlying hidden variables dynamics (DeBroglie-Bohm)...  Are we forced to include the detector into the quantum system, in order to compute transition probabilities, if the interaction between $S$ and $O$ is a gauge invariant interaction between gauge variables?  I think the answer is negative \cite{Rovelli:2007ep}, but I leave it open in this paper, which is confined to classical theory. }

\section{Conclusions}

The idea that gauge-variant quantities must be seen \emph{uniquely} as mathematical redundancy in the description of physics, deprived on any physical meaning, misses an  aspect of Nature. It refers to systems in isolation, disregarding the fact that  systems can couple via non-gauge-invariant quantities. Gauge variables are components of relational observables which depend on more than a single component. 

Gauge-variable quantities can be associated to measurement procedures, although procedures that lead to unpredictable numbers (``partial observables"). This is true in general relativity and Yang Mills theory, and in the general-covariant formulation of dynamics where  the time variable is an example of partial observable.

In particular, gauge-dependent quantities can be effectively used for describing the way a system affects an apparatus. Examples are: a material reference system for the diffeomorphism gauge; a local Lorentz frame in the tetrad formulation of general relativity; a reference fermion in the Yang-Mills internal space; and a clock, for the gauge associated to the hamiltonian constraint. The coupling to the apparatus is gauge invariant, but the measured quantity, viewed as a variable of the system alone, is not. 

The thesis of this paper is not that restricting to gauge-invariant observables is wrong. We can always enlarge a system to include any other coupled system and apparatus. (At least as long as we disregard quantum theory.  In quantum theory a notion of measurement might be unavoidable. According to many interpretations, if not all, a split of the world into ``system" and ``observer" is necessary to associate a value to physical quantities.)  The thesis of this paper is that restringing to gauge-invariant observables makes us blind to a fine structure of the world.

Gauge invariance is not just mathematical redundancy; it is an indication of the relational character of fundamental observables in physics. These do not refer to  properties of a single entity. They refer to relational properties between entities: relative velocity, relative localization, relative orientation in internal space, and so on.\footnote{There is also a weaker sense in which measures are relational: we measure a quantity by comparing it to some standard: that's why quantities are assigned units on some scale. This is not related to gauge invariance, although the idea of gauge originated from Weyl's considerations about the relativity of scale.}

Gauge interactions describe the world because Nature is described by relative quantities that refer to more than one object.   In a sense, this is a  step along the direction devised by Galileo, when he stressed that velocity is a quantity that does not refer to a single object, but to two objects: the velocity of an object is only defined in relation to another object. 

Gauge is ubiquitous. It is not unphysical redundancy of our mathematics. It reveals the relational structure of our world.

\vspace{4mm}\centerline{***} \vspace{4mm}
Thanks to Eugenio Bianchi and Hal Haggard for a critical discussion of the paper and suggestions, and to Alejandro Perez for a lifelong discussion on this issue.


\end{document}